\documentclass[10pt]{article}
\usepackage[T1]{fontenc}
\usepackage{hyperref}
\usepackage{graphicx}
\usepackage{amsmath}
\usepackage{amssymb}

\newcommand{\bb}{\begin{equation}}
\newcommand{\ee}{\end{equation}}
\newcommand{\bega}{\begin{eqnarray}}
\newcommand{\ega}{\end{eqnarray}}
\newcommand{\begae}{\begin{eqnarray*}}
\newcommand{\egae}{\end{eqnarray*}}

\newcommand{\h}{\hspace*{4ex}}

\newcommand{\cent}{\centerline}
\newcommand{\vs}{\vspace*}

\begin{document}

\baselineskip 0.5cm


\begin{center}

{\large {\bf Optical reconstruction of computer-generated holograms 3D scenes via spatial light modulators } } 

\footnotetext{$^{\: (\dag)}$ E-mail addresses for contacts: marcos.gesualdi@ufabc.edu.br}


\end{center}

\vs{0.2 cm}

\cent{Mauricio Nel Bolanos Pena and Marcos R.R. Gesualdi $^{\: (\dag)}$}

\vs{0.2 cm}

\centerline{{\em Universidade Federal do ABC (UFABC), Santo André, SP, Brazil.}}

\vs{0.5 cm}

{\bf Abstract  \ --} \ This work presents the optical reconstruction of computer-generated holograms (CGHs) 3D scenes via spatial light modulators (SLMs). Holography is an optical technique that allows the recording and reconstruction of images of 3D objects, as a hologram carries the intensity and phase information about the object.  With the evolution of data storage and processing in computers, the viability of generating holograms by computers has increased. Currently the experimental implementation of CGHs (Computer-Generated Holograms) is done through new optoelectronic devices such as SLMs (Spatial Light Modulators) and high-resolution CCDs cameras, in addition to new photosensitive materials, allowing the optical recording and reconstruction of 3D images and special optical beams.  Therefore, presented here with good results are CGHs numerical construction (recording) of images both static and dynamic, 2D and 3D, and its reconstruction (reproduction) using optical SLMs devices to obtain 3D images and static and dynamic scenes.












%

\section{Introduction} 

\noindent Holography is an interferometric-diffractive technique that records the light distribution field (amplitude and phase) spread over an object as opposed to a photographic image in which it only records the amplitude (intensity), which allows the optical reconstruction of the 3D image from objects. The principles of   holography were proposed by Dennis Gabor in 1948, based on wave phenomena of light (interference and diffraction) [1]-[2].  The amplitude and phase information about light scattered by an object is recording in an interference pattern of the reference and object wavefront being recorded on a recording medium results in a hologram. This hologram it is illuminated with the reference wavefront and it allows reconstructing the scene by the diffraction in the holographic grating.  This discovery of high quality laser holograms has thus attracted  the attention of scientists [3]-[4].

\h Currently, holographic techniques can be classified according to the registration process or reconstruction process, as in analogical holography (classical conventional) and computational holography. In analogical holography, using laser and traditional optical elements, the registration process and reconstruction is done using silver halide films, thermoplastics or photorefractive materials [3]. Moreover, computational holography and the registration process or reconstruction of the hologram are made with the aid of optoelectronic devices and computers for reconstruction of a hologram.

\h With progress in semiconductor-based technologies, and in increased data processing and storage capacity in informatics, improved procedures have been employed, paving the way to digital holography. With devices like CCDs (Charge Coupled Devices) cameras, and the SLMs (Spatial Light Modulators) that are based on liquid crystal or micro mirrors the phase intensity of the waves can be measured, stored, transmitted and used in computer-controlled simulations [5]-[8].

\h Thus, varied applications have brought about the arrival of the CGHs and have resulted in the production of diffractive optical elements that are computer generated with spatial filters for optical information processing, and a large amount of research development in 3D display systems [9]-[10]. Advances in computing combined with the development of new algorithms such as Fast Fourier Transform marked the beginning of Computer Generated Holograms [3]-[11].

\h Moreover, holographic interferometry techniques have led to development of diverse applications such as nondestructive testing (test strain tensions materials, microscopy, fluids and other types of analysis) [7]-[12]-[13]. Also CGHs have found other purposes, such as in the area of medicine with images from human body in 3D design perspectives to help improve diagnostic [14]-[15].  Similarly, CGHs provide a tool for the modeling studies of non-diffracting beams, with the objective implementation in the areas of optical impingement, medicine, remote sensing and others [16]-[17].  This work wants to show that the computer generated holography studies from 3D virtual scene algorithms (software), can perform the optical reconstruction of these holograms using SLMs devices.  Three algorithms used for the numerical generation of CGHs were studied. Where the mathematical models used in each routine are described and implemented via MATLAB, getting both static and dynamic CGHs. Furthermore, the methods are processed for reconstruction of CGHs both statically and dynamically. Implemented in this way was an experimental arrangement for the optical reconstruction of CGHs via SLMs, yielding good results of both 3D images (static) and 3D scene (dynamic) objects.

\section{Computer-Generated Holograms}

\noindent In CGHs, the waves are numerically stored in the hologram from computer numerical calculus algorithms (software) or via hardware. In both cases, the optical reconstruction of the recorded waves are obtained by the diffraction of the light scalar. Particularly, the CGHs are designed considering the diffraction scale of light, the optical characteristics of the medium in which the hologram will be produced which may have a distribution phase and amplitude, and the light distribution characteristics of the hologram reconstruction plane. We can also produce a CGH by the Fourier or Fresnel methods. A Fourier hologram uses Fraunhofer diffraction to calculate the spread of the hologram plane of light to the reconstruction plane. A Fresnel hologram employs the Fresnel diffraction or special filtering to propagating light by calculating the free space between the plane of the optical hologram reconstruction plane.

\h The construction of a CGH can be simple, considering the two levels of transmittance zero and one [11]. To produce such a hologram, this should be analyzed as a set of NxN cells corresponding to NxN transforming the Discrete Fourier coefficients in the object plane. Therefore, each complex Fourier component is represented by a simple transparent area within the corresponding cell, whose size is determined by the Fourier coefficient of the module, but their position within the cell is the Fourier coefficient used in this step. The displacement of the transparent area in any cell results in transmitted light observable by a long or short path to reconstruct the image [23].

\h An effective method in generating CGHs is the concept of light point, which is a numerical approach to the hologram recording area. In summary, the object in question is assumed to be as a composition of points of light, each contributing to the formation of a hologram area known as Fresnel. The three-dimensional hologram is created by adding the Fresnel areas corresponding to each individual point of the object's light points, this type of hologram is known as a Fresnel hologram.

\section{Theory and Algorithms of CGHs} 

\noindent The mathematical modeling of optical beams proposed herein describe the amplitude and phase distribution of the object and reference wavefront taken from the algorithms worked out in MATLAB software. This software simplifies the storage and processing of the scenes allowing the images to be dealt with as an array of N rows x M columns. Each array element corresponds to a pixel of the image and has a range value of 0 representing the color black to 255 which is shown with its peak value as white. For generation of these CGHs we studied three types of algorithms: the Ray Tracing Method, the Amplitude Hologram and Fourier Hologram.

\subsection{Ray Tracing Method} 

\noindent In this method light beams are reflected from each point of the image and are mathematically constructed as a diverging light source. Therefore, the equation used for the light distribution at each point in space is defined in (1), which carries the phase and amplitude object information.

      \begin{equation}
E(x,y,z)=Ae^{i2{\pi}r/\lambda}
\label{equ:1}
\end{equation}

\noindent The variable A represents the field amplitude of the image point while $r$ is the distance from the light spot to the recording plane (record) of the hologram, and $\lambda$ is the laser beam wavelength, and what is intended to work in the experimental phase hologram reconstruction. Through the Ray Tracing Method we are looking to create the interference pattern between the reference beam and the reflected rays (scattered) by the object or scene. This object will be represented in the xy plane with variables, parallel to the hologram plane $x'y'$ at a distance along the z axis. To calculate the possible values of A(x,y) used are 1 or 0.  For simplifying the analysis of the spokes it is assumed that the reference beam is incident perpendicularly to the plane of hologram recording. The distribution field hologram plane $x'y'$ is defined in (2) [19].

\begin{equation}
H(x',y')= \int_{-\alpha}^{+\alpha}\int_{-\alpha}^{+\alpha}  A(x,y)e^{ikr}dxdy
\label{equ:2}
\end{equation}

\noindent where 

\begin{center}
$r=\sqrt{(x-x')^2 + (y-y')^2 + z^2 }$ e $k=2\pi/\lambda$. \\
\end{center}

\h To create a holographic image reading procedure, it is stored in a variable type array, given this, the scene is taken to the xy plane as a set of points centered on the source.  The objects points included in the $xy$ plane, the parameters were established to create the hologram. Its main features are the space where the hologram will be recorded and the wavelength, taking into account the SLM properties screen and laser beam for optical reconstruction. The spacing of pixels on the $x'y'$ plane will be defined by the pixel size and the SLM screen resolution which will be defined by an engraved image.

\subsection{The Amplitude Hologram} 

\noindent If the hologram scale scene is defined as a square matrix, then the hologram recording plane will be defined by the size and number of pixels on the SLM screen. The hologram is obtained through a transmission function defined in (4) using the amplitude modulation comprising the varying of the transmittance or reflection of the medium.

		\begin{equation} \label{hologram_2}
				H(x,y)=1/2\left\{\beta(x,y) + \alpha(x,y)\cos[\phi(x,y)-2\pi(\xi x + \eta y)]\right\}
		\end{equation}

\noindent The information that brings the wave object, i.e. the complex field can be set by the amplitude $\alpha(x,y)$ and the phase function $\phi(x,y)$; $(\xi,\eta)$ is the special frequency of the plane wave used as a reference beam and $\beta(x,y)$ defined by (5), in a smooth amplitude cover $\alpha(x,y)$ to decrease the influence of the central spectrum. Therefore, the hologram interference pattern is created by the field $\phi(x,y)$ wave object with plane wave $\exp[i2\pi(\xi x + \eta y)$.

        \begin{equation} 
          \beta(x,y)=[1+\alpha^{2}(x,y)]/2
        \end{equation}

\subsection{Fourier Hologram}  

\noindent Using the Fraunhofer diffraction theory there is a modeled propagation of the wavefront to generate the hologram. But the object plane is represented by the variables $(\xi,\eta)$ and the plan of the hologram by the variables $(x,y)$, the functions of the object beam and references are $To(\xi,\eta)$ and $Tr(\xi,\eta)$ respectively. Therefore, the Fraunhofer Diffraction formula relates to the amplitude $U(x,y)$ in the far field (construction plane hologram) with object plane by the equation (6) [21]-[23].

        \begin{equation} 
            U(x,y) = 
						\frac{{e^{ikz}}{e^{i\frac{k(x^2 + y^2)}{2z}}}}{i \lambda z} \int^{-\infty}_{\infty}\int^{-\infty}_{\infty} T(\xi,\eta) e^{-i\frac{2\pi}{\lambda z} (x\xi +y\eta)} d\xi d\eta \; .
        \end{equation}

\h The interference pattern of the calculation is obtained by adding the wavefront of the object and the reference in the hologram plane. From the linearity of the Fourier transform theorem, the two wave fields (object and reference) are added, then equation (6) and thus achieving their calculation.

        \begin{equation} 
				   U(x,y) =
						\frac{{e^{ikz}}{e^{i\frac{k(x^2 + y^2)}{2z}}}}{i \lambda z} \int^{-\infty}_{\infty}\int^{-\infty}_{\infty} [T_{0}(\xi,\eta) + T(\xi,\eta)] e^{-i\frac{2\pi}{\lambda z} (x\xi +y\eta)} d\xi d\eta \; .
        \end{equation}

\section{Dynamics Hologram Generation}  

\noindent For the generation of dynamic scenes, the scene or object is represented in 3D form using design software such as CAD (Computer-Aided Design). Using the CAD successive images are taken of the rotating object, with the purpose of creating each hologram of a dynamic scene. There were three were created in 3D CAD: the first was of the UFABC logo, the second a gear wheel or pinion and, finally, an encapsulated integrated circuit, Figure 1. Having a set of successive images of figures in 3D, these are grouped into film form in an AVI video format (Audio Video Interleave), and thus created is a dynamic scene to then be inserted into the SLM for later optical rebuilding.

         \begin{figure}[!ht]
                \centering
                \includegraphics[scale=0.75]{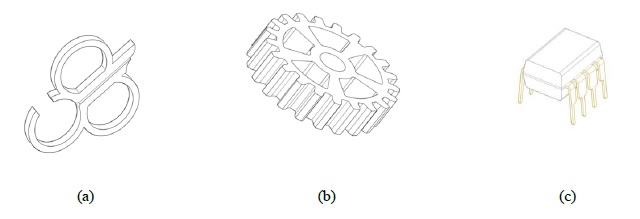}
                \caption{. 3D objects drawn in Autocad. (A) Federal University of ABC logo. (B) Pinions. (C) Integrated circuit based on the encapsulation of LM555.}
                \label{fig1}
        \end{figure}

         \begin{figure}[!ht]
                \centering
                \includegraphics[scale=0.75]{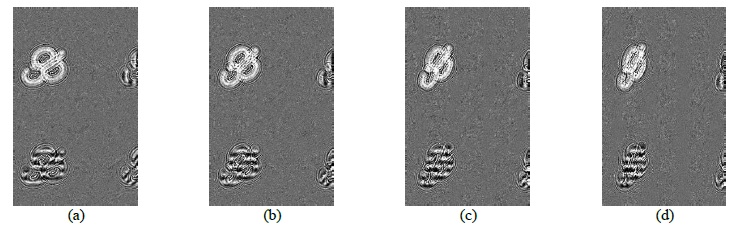}
                \caption{. Holograms sequence obtained from Fig. 1 (a) in order to create an object scene rotating shaft. Shown here are the first eight scenes 60 holograms generated.}
                \label{fig2}
        \end{figure}

         \begin{figure}[!ht]
                \centering
                \includegraphics[scale=0.75]{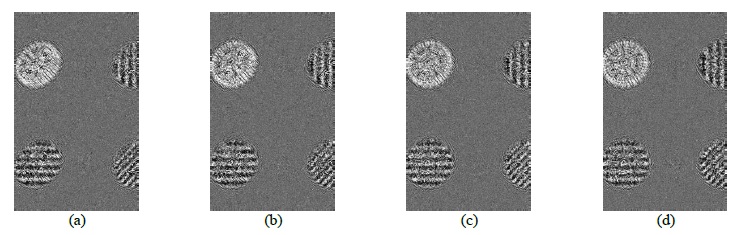}
                \caption{Sequence obtained holograms of Fig. 1 (b). The gear wheel is rotated in sequence for each hologram.}
                \label{fige3}
        \end{figure}

\

\h In Figures 2 and 3 are shown the sequence of holograms generated by the Ray Tracing Method. Objects in 3D perspectives were drawn in Autocad software which allowed the manipulation of these objects. The objects were carefully rotated on their axis at a constant angle, and in each turn, the view of the object was captured from a fixed perspective, and thus was achieved the number of necessaries frames, based on the acquisition rate and update of the CCD camera and SLM respectively. Then the images are edited in size and resolution by the GIMP editor software. Using a MATLAB routine for each CGH frame generated, a holographic sequence, i.e. a dynamic CGH was created. Figures 4, and 6 shows this sequence of CGH created by the Amplitude Hologram algorithm.

		    \begin{figure}[!ht]
                \centering
                \includegraphics[scale=0.75]{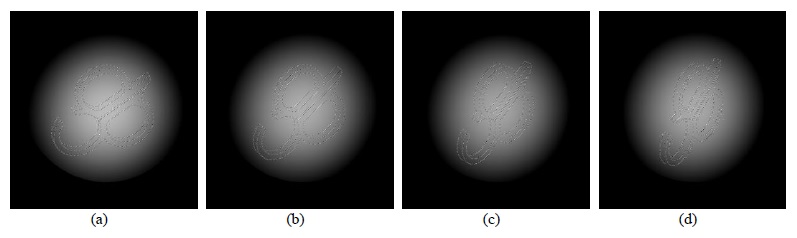}
                \caption{Hologram following the UFABC logo through Amplitude Hologram algorithm, the sequence with an envelope in order to take the order component to zero.}
                \label{fig4}
        \end{figure}

    \begin{figure}[!ht]
                \centering
                \includegraphics[scale=0.75]{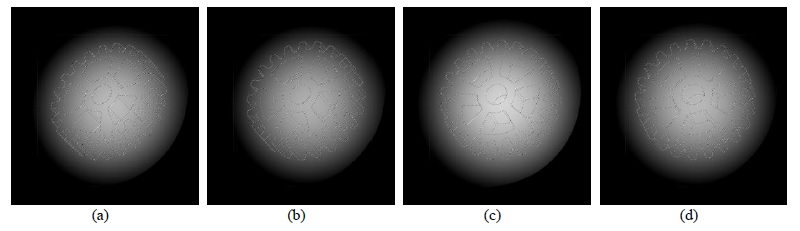}
                \caption{Hologram following the gear wheel with the Amplitude Hologram algorithm. The sequence shows the smooth cover to reduce the effect of the zero order, it is necessary to generate a good holographic reconstruction.}
                \label{fig5}
        \end{figure}

\begin{figure}[!ht]
                \centering
                \includegraphics[scale=0.75]{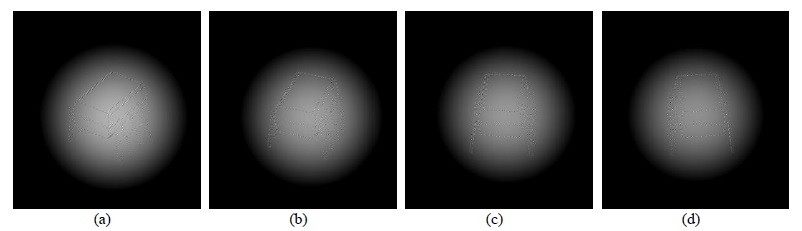}
                \caption{Holographic sequence of an Integrated Circuit with the Amplitude Hologram algorithm. The sequence shows the envelope for reducing the effect of the zero order, necessary to generating a good reconstruction.}
                \label{fig6}
        \end{figure}

\

\section{Experimental Holographic Reconstruction of Static and Dynamics CGHs via SLMs}

\noindent AThrough the optical reconstruction of CGHs obtained in MATLAB is it possible to check the quality of the subjects algorithms. For the needed reconstruction an optical image processing system with special filtering 4f composed of two lenses with the same focus. The wave propagating through the system is to be recognized as a cascading of two Fourier transformed subsystems: the first subsystem between the object plane and the Fourier plane performs a Fourier transform; and the other subsystem between the Fourier plane and image by an inverse Fourier transform.

    \begin{figure}[!ht]
                \centering
                \includegraphics[scale=0.75]{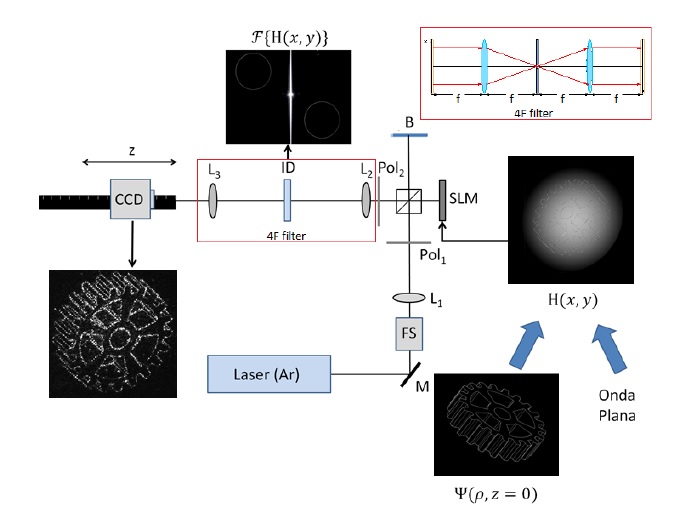}
                \caption{Holographic setup, where Fe1 is a spatial filter and lens Le1 helps define the intensity and size of the circular area of the laser beam; Es is the mirror that directs the beam; Po1 and Po2 are polarizing where the first controls the intensity of light and the second the level of noise; DF splits the beam; SLM will contain the hologram (CGH) that is sent from the PC computer; LE2 and LE3 is part of the optical system 4f; and CCD camera will make the image capture rebuilt by the system. In detail, the optical reconstruction system with filtering 4f, where Lens 1 makes the Fourier transform, the opening makes the selection of the frequential component and the lens 2 makes the inverse Fourier transform.}
                \label{fig7}
        \end{figure}

\

\noindent In order to achieve the optical reconstruction of generated holograms (CGHs) with the algorithms in MATLAB, a holographic optical setup is made (Figure 7), and an optical reconstruction system CGHs via SLMs is implemented. The experimental arrangement was mounted on an optical table with a  damping system; a He-Ne laser and 633nm 17MW, optical and mechanical systems, a monochrome CCD camera with pixel size 4,65x4,65$\mu m$m and up to 15 FPS and an LC display reflective SLM recording rate with a resolution of 1920x1200 with a pixel size of 8,1$\mu m$.

\h The laser beam or light source used will go through the system and be treated in order to obtain a flat polarized wave, which focuses on the SLM screen on which are loaded the data resulting from CGHs algorithms worked out in MATLAB. The SLM devices work in a way to control the transmission (or reflection) of the light using a control of the polarization of light, where each level of polarization corresponds to a voltage applied in each cell (pixel). The CGHs are generated in an 8-bit image system so that it has 256 levels of gray (black is zero and white is 255). Thus, the polarization control has the same number of gray levels in controlling the transmission (or reflection) of the light. The wave is reflected by the screen of the SLM and is guided by the system 4f beam splitter, by which, in the first lens out, having a plane of  frequency spatial components, a 4F filter is used to make frequency selection in the Fourier plane (Fourier domain). As the L1 lens separates the frequency orders in the Fourier plane (lens focus), using a mask the first frequency order containing the holographic image information [11] is selected. At the exit of the 4f system, the second lens regroups the components by optical reconstruction, where a CCD camera will capture the reconstructed image.

\h For the reconstruction of dynamic CGHs experimental arrangement of Figure 7 a dynamic hologram is inserted through a multimedia display software such as VLC Media Player or an implay function of MATLAB, while the SLM is characterized as a monitor on which it will be viewed on holographic scene dynamics. When implementing dynamic CGH in 4f experimental setup in multimedia software, in the case of VLC Media Player there are some disadvantages, such as not being able to reduce the increase in the size of the dynamic scene, and thus take advantage of the SLM screen space. For other hand, with the implay function MATLAB offers the option to specify the size of the scene. Besides the multimedia software there was implemented a remote Teamviewer monitoring software for handling the visual setting CGH in SLM. The CGH is being played though multimedia software and is taken to SLM through Teamviewer, which through a polarized wave that focuses on the SLM screen is reflected leading to SLM information and to the first lens filtration system 4f, getting the first dynamic diffraction orders in the Fourier plane. The order with which the chosen object information is brought to the second lens system to be rebuilt and captured by the CCD camera. Thus recorded is the dynamic and reconstructed scene through the arrangement shown in Figure 7 (in detail).

\subsection{Reconstruction Holograms Static Fourier Algorithm}

\noindent The Fourier algorithm was used with the images in Figure 8.  For the calculation of the Fourier hologram, a random phase was considered and added to the matrix form of the object, allowing for diffraction order spacings, facilitating the choice of the Fourier component that has the information on the wave object and thus optimizing the reconstruction of the hologram.

        \begin{figure}[!ht]
                \centering
                \includegraphics[scale=0.75]{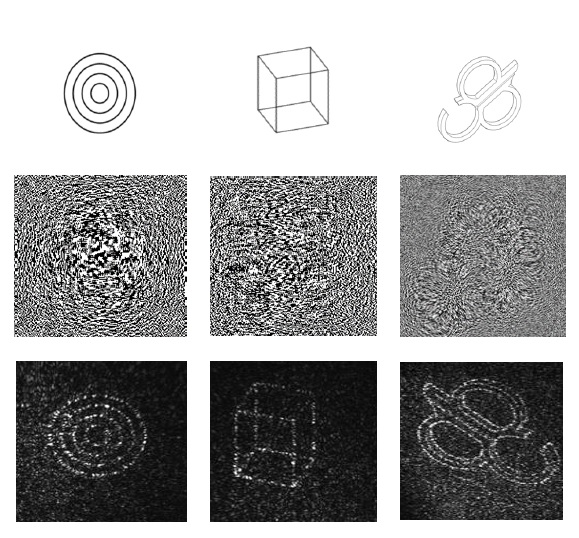}
                \caption{Images processed by Fourier algorithm, concentric rings, isometrics and soon the UFABC. (a) Hologram Fourier algorithm from the image of concentric rings, the isometric image and UFABC logo. And, Optical image reconstruction of concentric rings, the isometric image and UFABC logo.}
                \label{fig8}
        \end{figure}

\h At the time of optical reconstruction there are distinguishable diffraction order differences in the Fourier plane, due in part to the effect of spacing between the cells of the network pixels of the SLM screen, making it difficult to filter the diffraction order containing information about the registered scene in CGH.

\subsection{Reconstruction Holograms by Dynamic Ray Tracing Method}

\noindent The reconstructions of dynamic CGHs presented here were generated from the Rays Tracing Method. In Figure 9 it is shown that the dynamic reconstruction of scenes captured by the CCD camera captured the UFABC logo and the gear wheel. Despite the fact that the reconstructions show a great deal of information regarding  the object wavefront, that is, one can see the shape of the object with clarity even though the film has certain interference. This is due to low diffraction efficiency generated by the algorithm employed in generating the CGH since the object and references wavefronts fall perpendicularly to the hologram plane (on-axis size). However, the different diffraction orders are aligned on the optical axis, propagating in the same direction, and at the time of reconstruction can be seen at the same time [20]-[23].

\begin{figure}[!ht]
                \centering
                \includegraphics[scale=0.75]{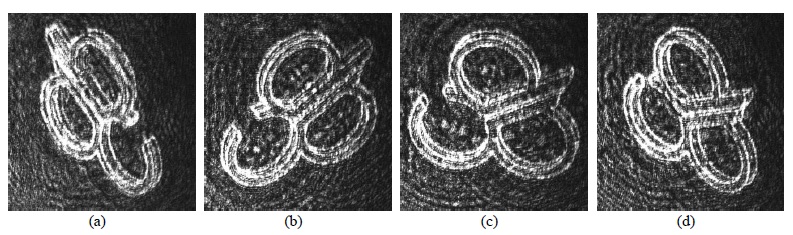}
                \caption{Dynamic sequence of optical reconstruction of the UFABC logo, generated with the Ray Tracing Method.}
                \label{fig9}
        \end{figure}

\begin{figure}[!ht]
                \centering
                \includegraphics[scale=0.75]{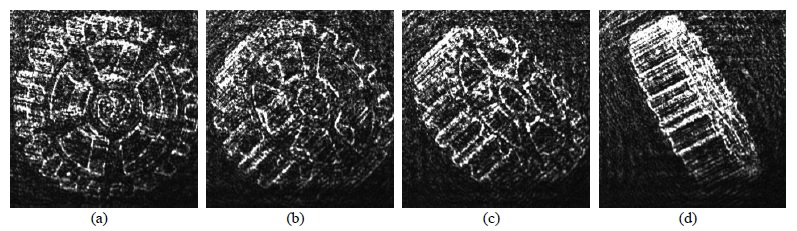}
                \caption{Dynamic sequence of optical reconstruction of the gear wheel, generated with the Ray Tracing Method.}
                \label{fig10}
        \end{figure}

\

\subsection{Reconstruction of Holograms by Dynamic Amplitude Algorithm}

\noindent The reconstructions obtained with the Amplitude Hologram feature scenes with the lowest level of interference. However, the reconstruction of dynamic amplitude holograms has higher diffraction efficiency. The algorithm used in the transmission function considered as a function of $\beta(x,y)$ which is a smooth amplitude cover to reduce the influence of the central spectrum. In the smooth cover, the front of the mathematical waveform contains a reference angle of incidence $\theta$ to the optical axis, preventing the different diffraction orders that they overlap. 

\begin{figure}[!ht]
                \centering
                \includegraphics[scale=0.75]{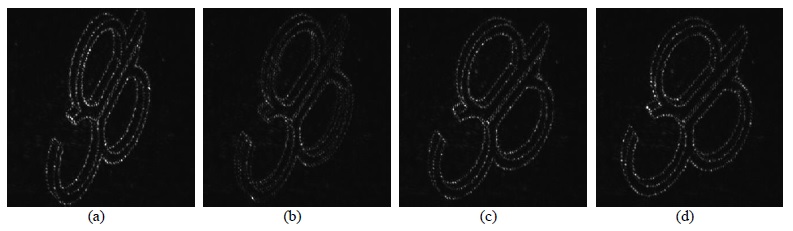}
                \caption{Following the dynamic optical reconstruction of the UFABC logo, generated from the Amplitude Hologram algorithm.}
                \label{fig11}
        \end{figure}

\

\begin{figure}[!ht]
                \centering
                \includegraphics[scale=0.75]{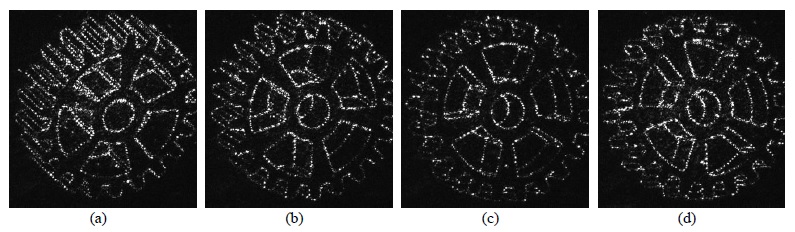}
                \caption{Dynamic sequence of optical reconstruction of the gear wheel, generated from the Amplitude Hologram algorithm.}
                \label{fig12}
        \end{figure}

\

\begin{figure}[!ht]
                \centering
                \includegraphics[scale=0.75]{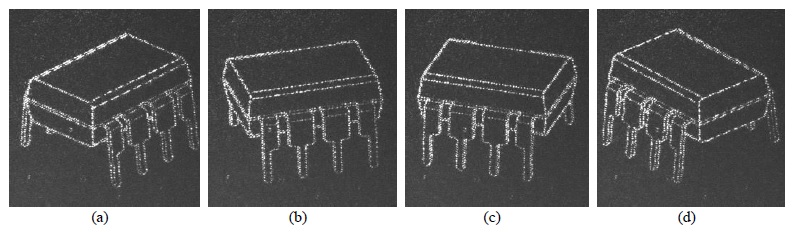}
                \caption{.  Following the dynamic optical reconstruction of the encapsulated integrated circuit, generated from the Amplitude Hologram algorithm.}
                \label{fig13}
        \end{figure}

\

\section{Conclusions}

\noindent In this work we have shown an optical reconstruction system of computer generated holograms 3D scenes via SLMs using various algorithms and CGH generation via MATLAB software.  Through the environment contained in MATLAB, both static and dynamic CGHs scenes were generated through 2D and 3D perspectives, with the help of implementation of algorithms such as: Ray Tracing Method, Fourier Hologram and Amplitude Hologram. 

\noindent In  the CGHs obtained by the Lightning Stroke method, in spite of presenting an overlap in different diffraction orders, a great amount of the information record was saved  of the scene, as much as in the static reconstructions, as in  the dynamic reconstructions. It should be noted that this process can be improved, by applying a random phase to each object point or a constant phase the reference beam.

\noindent In the algorithms used for the Fourier Amplitude, the processing time is much shorter compared to the Ray Tracing Method.  For the mathematical model of the Fourier Amplitude Hologram, the object and reference beam is matrixed so that the matrix with calculus are mathematical functions contained in MATLAB. Unlike the Ray Tracing Method where the data is arranged in a vector form  and its process has a larger number of iterations.

\noindent In CGH Amplitude Hologram, the static and dynamic holograms show greater efficiency diffraction. The reconstructed wavefronts have lost less information in proportion to the Ray Tracing Method and CGH Fourier algorithm. This is due in part to the different diffraction orders that are not overlapping and  the optimization of the selection process of the diffraction orders containing information.

\noindent The algorithms for the generation of static and dynamic computer produced holograms (Ray Tracing Method, Fourier Hologram and Amplitude Hologram), contribute greatly to different types of research being done in the areas of photonics such as, optical reconstruction of 3D images and non-diffracting beams and/or as a tool in the modeling of micro masks for manufacturing semiconductor devices by photolithographic process used for material awareness.

\

\section{Acknowledgments}

\noindent This research is supported by the UFABC, FAPESP (grant 09/11429-2) and CNPQ (grants 476805/2012-0 and 313153/2014-0).

\ 

\newpage



\end{document}